\documentclass[paper,12pt]{JHEP3}
\usepackage[centertags]{amsmath}
\usepackage{amsfonts} \usepackage{amssymb} \usepackage{amsthm}
\usepackage{bbm}
\usepackage{bm}

\def\ie{{\it i.e.}~}

\def\ie{{\it i.e.\ }}
\newcommand{\be}{\begin{equation}}
\newcommand{\ee}{\end{equation}}
\newcommand{\bea}{\begin{eqnarray}}
\newcommand{\eea}{\end{eqnarray}}
\newcommand{\ba}{\begin{array}}
\newcommand{\ea}{\end{array}}

\title{Recursion relations in CFT and N=2 SYM theory}
\author{\parbox{11.5cm}{Rubik Poghossian}\\
\vspace{0.3cm}

I.N.F.N.,  Universit\`a di Roma Tor Vergata\\
Via della Ricerca Scientifica, I-00133 Roma, Italy\\
and Yerevan Physics Institute, Alikhanian Br. 2\\
0036 Yerevan, Armenia\\
\email{poghos@yerphi.am}\\
\email{poghosyan@roma2.infn.it}
}
\abstract{
Based on prototypical example of Al.Zamolodchikov's recursion relations for the four point conformal block 
and using recently proposed Alday-Gaiotto-Tachikawa (AGT) conjecture, recursion relations are derived for the generalized
prepotential of ${\cal N}=2$ SYM with $f=0,1,2,3,4$ (anti) fundamental or an adjoint hypermultiplets. In all cases 
the large expectation value limit is derived explicitly. A precise relationship between generic 1-point conformal 
block on torus and specific 4-point conformal block on sphere is established. In view of AGT conjecture this translates 
into a relation between partition functions with an adjoint and 4 fundamental hypermultiplets.     
}
\preprint{ROM2F/2009/17}
\begin{document}
\section{Introduction}
Recently Alday Gaotto and Tachikawa \cite{AGT} have found a very remarkable relation between generalized partition 
functions \cite{Nek,FP,BFMT,NekOkoun} in certain classes 
of ${\cal N}=2$ conformal $SU(2)$ quiver gauge theories \cite{Gaiottodualities} (see \cite{FMPquivers} for instanton counting in quiver theories) 
and correlation functions of 2d Liouville theory on Riemann surfaces. In this paper I will consider only the "holomorphic" version  
of AGT conjecture concerning the relation between instanton part of the partition function in gauge theory and the conformal 
block in 2d CFT. More precisely I'll concentrate on $SU(2)$ gauge theories with four fundamental hypermultiplets and the theory 
with an adjoint hypermultiplet. This choice is of particular interest since they are related to such fundamental objects of 2d CFT 
as 4-point and torus 1-point conformal blocks. I will adopt in this paper a slightly generalized version of the original AGT conjecture and will not 
assume a specific relation between Nekrasov's deformation parameters $\epsilon_{1}$ and $\epsilon_{2}$, a point 
emphasized also in recent works \cite{MarMirMor1,MarMirMor2}. Another deviation from the initial AGT will be discussion
 of non-conformal gauge theories (a possibility also discussed in very recent papers \cite{Gaiottononconf, MarMirMor2} 
 from a different perspective).     
 
In all further discussions the recursion relation for the CFT 4-point conformal 
block discovered by Alexei Zamolodchikov \cite{AlZam} a quarter of century ago will play the central role. Unfortunately this brilliant work 
is not widely known even by the specialists.

The section \ref{qrecursion}. is a breaf introduction to the Zamolodchikov's recursion relation.  

The section \ref{Zinst}. is devoted to the 
description of the instanton part of the generalized partition function in ${\cal N}=2$ SYM theories introdced by Nekrasov 
\cite{Nek}.  Representation of the Nekrasov partition function as a sum over (multiple) Young tableau 
in a way suitable for practical higher order instanton calculations is based on the character formula describing decomposition of the tangent 
space of the moduli space of instantons under the combined (global gauge plus space-time rotations) torus action around fixed 
points \cite{FP}. The Nekrasov partition function for the cases with different types of extra hypermultiplets 
can be read of from the character formula incorporating specific factors which depend on the representations of the hypermultiplets 
\cite{Nek,BFMT,NekOkoun}. In this section for further reference the cases of fundamental or adjoint hypermultiplets are presented 
in some details.

In section \ref{fundamentals}. the Zamolodcikov recursion relation for 4-point conformal block is translated into the relation 
for the partition function with four fundamentals. As a particular application the exact large vev asymptotic 
of the partition function is derived and the leading term is checked against the Seiberg-Witten curve \cite{SW} analysis. 
Considering the large mass limit when one or more fundamentals decouple the analogous relations for less number 
of hypermultiplets are derived. The recursion relation for the case without extra matter is especially simple and may 
serve as a convenient starting point for investigation of the analytical properties of the prepotential in presence 
of nonzero gravitational background. 

In section \ref{adjoint}. a similar recursion relation is proposed for the case of adjoint hypermultiplet. The conjectured 
relation has been checked against explicit instanton calculations up to order 5. Again the AGT conjecture leads to 
analogous (previously unknown in CFT) relation for the torus 1-point conformal block. Comparison of the results of 
this sections with those of previous one leads to a surprising conclusion:  the torus 1-point block is closely related to 
specific sphere 4-point block or alternatively the generalized ${\cal N}=2$ SYM partition function with adjoint hypermultiplet 
is related to the partition function with four hypermultiplets. 

Finally the appendix \ref{torusblock}. briefly describes how to calculate the torus one point conformal 
block from CFT first principles.

\section{Zamolodchikov's q-recursion relation for CFT conformal block}\label{qrecursion}

Though there is no closed analytic expression for the general 4-point conformal block, Al.~Zamolodchikov has found an extremely convenient 
recursion (Russian doll type) relation, which allows to calculate the conformal block up to the desired order in x-expansion. Below I give a brief 
description of Zamolodchikov's recursion relation closely following to \cite{ZamZamLiouville}. It is convenient to represent the generic $4$-point 
conformal block ${\cal F}$ as \cite{AlZam}
\bea
{\cal F}(\Delta_i,\Delta,x)=\left(16q\right)^{-\alpha^2}x^{Q^2/4-\Delta_1-\Delta_2}(1-x)^{Q^2/4-\Delta_1-\Delta_3} \nonumber\\
\times\theta_3(q)^{3Q^2-4(\Delta_1+\Delta_2+\Delta_3+\Delta_4)}H(\mu_i,\Delta,q)\label{H}
\eea  
where
\bea
\theta_3(q)=\sum_{n=-\infty}^\infty q^{n^2}
\eea 
Here $\Delta_i$, $i=1,2,3,4$ - the dimensions of the external (primary) fields (placed at the points $x$, $0$, $1$ and $\infty$ respectively) and $\Delta$ - 
the internal dimension are parametrized by
\bea
\Delta_i=\frac{Q^2}{4}-\lambda_i^2\,,\hspace{1cm} \Delta=\frac{Q^2}{4}-\alpha^2\,,\label{dimensions}
\eea 
where $Q$ is related to the central charge of the Virasoro algebra through
\bea
c=1-6 Q^2
\eea
For further purposes I parametrise the background charge $Q$ via
\bea
Q=\frac{\epsilon_1+\epsilon_2}{\sqrt{\epsilon_1\epsilon_2}}.
\eea
and introduce the parameters 
$\mu_i$ (later to be related with the masses of the anti-fundamental hyper-multiplets under the
AGT conjecture) as linear combinations of $\lambda_i$:
\bea
\mu_1&=&\lambda_1+\lambda_2+\frac{Q}{2}\,,\,\,\mu_2=\lambda_1-\lambda_2+\frac{Q}{2}\,,\,\,\nonumber\\
\mu_3&=&\lambda_3+\lambda_4+\frac{Q}{2}\,,\,\,\mu_4=\lambda_3-\lambda_4+\frac{Q}{2}\,.\label{mu-lambda}
\eea
Comparing with the standard $Q=b+1/b$ we see that $b=\sqrt{\epsilon_1/\epsilon_2}$.\footnote{ In fact upon simple rescaling 
of the masses and vev's of a ${\cal N}=2$ conformal SYM theory  by the factor 
$1/\sqrt{\epsilon_1\epsilon_2}$, all expressions become homogeneous in $\epsilon_{1,2}$. Thus there is no need 
to follow \cite{AGT} and restrict oneself to the case $\epsilon_1\epsilon_2=1$.} 
The parameter $q=e^{i\pi\tau}$ is related to the coordinate $x$:
\bea
\tau&=&i\frac{K(1-x)}{K(x)},
\eea
where 
\bea
K(x)&=&\frac{1}{2}\int_0^1\frac{dt}{\sqrt{t(1-t)(1-xt)}}\,\,.\label{tau}
\eea 
Conversely
\bea
x=\frac{\theta_2^4(q)}{\theta_3^4(q)},\label{xversusq}
\eea
where
\bea
\theta_2(q)=\sum_{n=-\infty}^\infty q^{(n+1/2)^2}.
\eea 
Here are the first few terms of the small $x$ expansion of $q$:
\bea
16 q=x+\frac{x^2}{2}+\frac{21 x^3}{64}+\frac{31 x^4}{128}+{\cal O}(x^5)\label{qexpansion}
\eea 
The asymptotic behaviour of the conformal block at large internal dimension $\Delta \rightarrow\infty$ has been established in \cite{AlZam} 
which in terms of the function $H$ is very simple: 
\bea
H=1+{\cal O}(\Delta)\label{asymptotic}
\eea  
Now we are ready to state Al.Zamolodchikov's q-recursion relation:
\bea
H(\mu_i,\Delta,q)=1+\sum_{m,n=1}^\infty\frac{q^{mn}R_{m,n}^{(4)}}{\Delta-\Delta_{m,n}} 
H(\mu_i,\Delta_{m,n}+mn,q), \label{AlZamrecursion}
\eea
where the poles are located at
\bea
\Delta_{m,n}=\frac{Q^2}{4}-\lambda_{m,n}^2
\eea
with
\bea
\lambda_{m,n}=\frac{m\epsilon_1+n\epsilon_2}{2\sqrt{\epsilon_1\epsilon_2}}
\eea
\ie exactly at the degenerated internal dimensions. Finally 
\bea
R_{m,n}^{(f)}=\frac{2\prod_{r,s}\prod_{i=1}^f(\mu_i-\frac{Q}{2}-\lambda_{r,s})}{\prod^\prime_{k,l}\lambda_{k,l}},\label{Rmn}
\eea
where the products are over the pairs $(r,s)$ and $(k,l)$ within the range
\bea
r&=&-m+1,-m+3,\ldots ,m-1\nonumber\\
s&=&-n+1,-n+3,\ldots, n-1\nonumber\\
k&=&-m+1,-m+2,\ldots, m-1,m\nonumber\\
l&=&-n+1,-n+2,\ldots, n-1,n\nonumber
\eea
while prime over the product in the denominator indicates that the pairs $(m,n)$ and $(0,0)$ should be suppressed.   

Note that the function $H$ in order to satisfy (\ref{AlZamrecursion}) should be totally symmetric w.r.t. permutations of $\mu_i$'s.  
Less obvious is the symmetry with respect to reflections $\mu_i\rightarrow Q-\mu_i$ accompanied with $q\rightarrow -q$ which is
a consequence of $\lambda_{-r,-s}=-\lambda_{r,s}$. There is no doubt that these remarkably reach symmetries 
of the 4-point block and their consequences are worth to be explored in greater details. 

In order to give some flavor how the recursion relation  (\ref{AlZamrecursion}) works in practice I give the result of iteration up to the
order $q^2$
\bea
 H=1+\frac{R^{(f)} _{1,1} q}{\Delta -\Delta _{1,1}}+\left(\frac{(R^{(f)} _{1,1})^2}{\Delta -
 \Delta _{1,1}}+\frac{R^{(f)} _{1,2}}{\Delta -\Delta _{1,2}}+\frac{R^{(f)} _{2,1}}{\Delta -\Delta _{2,1}}\right) q^2+{\cal O}(q^3)\label{Hexpansion}
\eea
Another elementary but useful observation is that at the order $q^l$ one encounters poles at $\Delta=\Delta_{n,m}$ with $nm<l$.

\section{Generalized partition function of ${\cal N}=2$ SYM with fundamental or adjoint hyper-multiplets}\label{Zinst}
In the seminal paper \cite{Nek} Nekrasov has proposed to generalize the Seiberg-Witten prepotential including into the game besides unbroken 
gauge transformation also the space time rotations which allowed to localize instanton contributions around finite number of fixed points. 
The general problem of computing the contribution of a given fixed 
point has found its final solution in \cite{FP}.  When the gauge group is $U(N)$ 
the fixed points are in
one to one correspondence with the arrays of Young tableau 
$\vec{Y}=(Y_1,...,Y_N)$ with total number of boxes $|\vec{Y}|$
being equal to the instanton charge $k$. The (holomorphic) tangent space
of the moduli space of instantons decomposes into sum of (complex) one dimensional 
irreducible representations of the Cartan subgroup of $U(N)\times O(4)$ 
\cite{FP} 
\bea 
\chi=\sum_{\alpha ,\beta =1}^N e_\beta e_\alpha^{-1} \left\{\sum_{s\in
Y_\alpha}\left(T_1^{-l_{Y_\beta}(s)}T_2^{a_{Y_\alpha}(s)+1}\right)+\sum_{s\in
Y_\beta}\left(T_1^{l_{Y_\alpha}(s)+1}T_2^{-a_{Y_\beta}(s)}\right)\right\},\label{char}
\eea 
where $(e_1,...,e_N)=(e^{ia_1}, \ldots, e^{ia_N})\in U(1)^N\subset U(N)$ and 
$(T_1, T_2)=(e^{i\epsilon_1},e^{i \epsilon_2})\in U(1)^2\subset O(4)$ , 
$a_{Y}(s)$ ($l_{Y_\alpha}(s)$) is the distance of the right edge of the box $s$ from the 
limiting polygonal curve of the Young tableaux $Y$ in horizontal (vertical) direction taken with
the sign plus if the box $s\in Y_\alpha$ and with the sign minus otherwise.

One-dimensional  subgroups of the above mentioned 
$N+2$ dimensional torus are generated by the vector fields parametrized by $a_1,\ldots,a_N$ and $\epsilon_1,\epsilon_2$. 
From the physical point of view $a_\alpha$ are the vacuum expectation values of the 
complex scalar of the ${\cal N}=2$ gauge multiplet and $\epsilon_1, \epsilon_2$ specify a particular 
gravitational background now commonly called $\Omega$-background.   
The contribution of a fixed point to the Nekrasov partition function in the basic ${\cal N}=2$ case without extra
hypermultiplets is simply the inverse determinant of the above mentioned vector field action 
on the tangent space at given fixed point. All the eigenvalues of this vector field can be directly read off from 
the character formula (\ref{char}) . The result is \cite{FP}   
\bea
P_{gauge}(\vec {Y})=\prod_{\alpha,\beta =1}^N\prod_{s\in Y_\alpha}\frac{1}{E_{\alpha ,\beta}(s)(\epsilon-E_{\alpha ,\beta}(s))},
\label{gauge}
\eea
where 
\bea
E_{\alpha ,\beta}=a_\beta-a_\alpha -\epsilon_1l_{Y_\beta}(s)+\epsilon_2(a_{Y_\alpha}(s)+1)
\eea
In general the theory may include "matter" hypermultiplets in various representations of the gauge group. 
In that case one should multiply the gauge multiplet contribution (\ref{gauge}) by another factor $P_{matter}$. 
In this paper we will consider the case of several (up to four) hypermultiplets in anti-fundamental representation 
and also the theory with an  adjoint hypermultiplet (so called ${\cal N}=2^*$). The respective matter factors read
 \cite{BFMT}
\bea
&&P_{antifund}(\vec {Y})=\prod_{l=1}^f\prod_{\alpha=1}^N\prod_{s_\alpha\in Y_\alpha}(\chi_{\alpha,s_\alpha}+m_l)\label{antifund} \\
&&P_{adj}(\vec {Y})=\prod_{\alpha,\beta =1}^N\prod_{s\in Y_\alpha}(E_{\alpha ,\beta}(s)-M)(\epsilon-E_{\alpha ,\beta}(s)-M),
\label{adj}
\eea
where $m_l$, $M$ are the masses of the hypermultiplets, $\epsilon=\epsilon_1+\epsilon_2$, 
\bea
\chi_{\alpha,s_\alpha}=a_\alpha+(i_{s_\alpha}-1)\epsilon_1+(j_{s_\alpha}-1)\epsilon_2
\eea
and  $i_{s_\alpha}$, $j_{s_\alpha}$ are the numbers of the column and the row of the tableaux $Y_\alpha$ 
where the box $s_\alpha$ is located.
Note also that in order to get a fundamental  hypermultiplet instead of an antifundamental, one should simply replace  
the respective mass $m_l$ by $\epsilon-m_l$ in (\ref{antifund}).
In terms of above defined quantities the instanton part of Nekrasov partition function reads\footnote{I use notation $x=e^{2\pi i\tau_g}$ 
with $\tau_g$ the usual gauge theory coupling to avoid confusion with the already introduced in chapter 2 parameter $q$ 
and to make comparison with 2d CFT block transparent.}
\bea
Z_{inst}=\sum_{\vec{Y}}x^{|\vec{Y}|}P_{gauge}(\vec {Y})P_{matter}(\vec {Y}) \label{zinst}
\eea


\section{AGT conjecture for the four-point conformal block and recursion relations for  ${\cal N}=2$ SYM with extra fundamentals}
\label{fundamentals}
From now on we will consider only the gauge group $SU(2)$ and will set the vacuum expectation values $a_1=-a_2=a$.

\subsection{f=4 antifundamentals}

According to the AGT conjecture for the case of $f=4$ extra antifundamental hypermultiplets one has \cite{AGT} 
\bea
Z_{inst}^{(4)}(a,m_i,x)=x^{\Delta_1+\Delta_2-\Delta} (1-x)^{2(\lambda_1+
\frac{Q}{2})(\lambda_3+\frac{Q}{2})}{\cal F}(\Delta, \Delta_i,x),\label{ZversusF}
\eea 
where $Z_{inst}^{(4)}(a,m_i,x)$ is given by (\ref{antifund}), (\ref{zinst}) specialized to the case of the gauge  group 
$SU(2)$ and  $f=4$ flavours. The vev $a=\alpha \sqrt{\epsilon_1\epsilon_2}$ and masses 
$m_i=\mu_i \sqrt{\epsilon_1\epsilon_2}$ are related to the conformal dimensions 
$\Delta$, $\Delta_i$ through  (\ref{dimensions}), (\ref{mu-lambda}). Equivalently, taking into account (\ref{H}):
\bea
Z_{inst}^{(4)}(a,m_i,x)&=&\left(\frac{x}{16q}\right)^{\alpha^2}(1-x)^{\frac{1}{4}(Q-\sum_{i=1}^4\mu_i)^2} \nonumber\\
&&\times\left[\theta_3(q)\right]^{2\sum_{i=1}^4(\mu_i^2-Q\mu_i)+Q^2}H(\mu_i,\Delta,q).\label{ZversusH}
\eea

Thus Zamolodchikov's recursion relation (\ref{AlZamrecursion}) for four point conformal block automatically 
provides a very efficient tool also for calculating the partition function  of the $SU(2)$ SYM theory with extra 
four (anti) fundamental hypermultiplets. One obvious advantage of the recursion relation compared to the explicit 
formula (\ref{zinst}) is that at each order of the "renormalized" (through the relation (\ref{xversusq})) 
instanton parameter $q$ the former immediately  determines the pole structure in variable 
$\Delta=Q^2/4-\alpha^2$ (see remark after Eq. (\ref{Hexpansion})).
Alternatively the formula (\ref{zinst}) is a sum over all (rapidly growing number of) couples of Young tableau each term being 
a simple factorized rational expression. Unfortunately the poles of the individual terms are extremely redundant: most of the poles 
 after summation of all terms of given instanton order disappear. In this sense the Zamolodchikov recursion relation and 
the explicit formula (\ref{zinst}) play complementary roles: the recursion relation provides a powerful tool for investigation 
of the analytical properties of the Nekrasov partition function while the Eq. (\ref{zinst}) together with Eq. (\ref{ZversusH}) 
provide a closed expression for the four point conformal block. Needless to say both tasks are of considerable importance 
and were waiting long time to find a solution. 

As a  most immediate consequence of Eq. (\ref{ZversusH}) one learns the asymptotic behaviour of the partition function at 
large vev's 
\bea
Z_{inst}^{(4)}(a,m_i,x)\sim\left(\frac{x}{16q}\right)^{\frac{a^2}{\epsilon_1\epsilon_2}}
(1-x)^{\frac{1}{4 \epsilon _1\epsilon _2}\left(\epsilon-\sum _{i=1}^4 m _i\right)^2} \nonumber\\
\times\left[\theta_3(q)\right]^{\frac{1}{\epsilon _1\epsilon _2}\sum _{i=1}^4 \left(m _i^2+
\left(\epsilon-m _i\right)^2-
3 \epsilon^2/4\right)}.\,\,\,\,\,\,
\eea
The instanton part of the Seiberg-Witten prepotential is
\bea
F^{SW}_{inst}=-\lim_{\epsilon_{1,2}\rightarrow 0} \epsilon_1 \epsilon_2Z_{inst}
\eea
hence
\bea
F^{SW}_{inst}\sim a^2\log\frac{16q}{x}-\frac{1}{4}\left(\sum_{i=1}^4m_i\right)^2\log(1-x)-2 \sum_{i=1}^4m_i^2 \log  
\theta_3(q)\
\eea
Taking into account Eq. (\ref{xversusq}), (\ref{qexpansion}) for the first leading in $a^2$ term one gets 
\bea
F^{SW}_{inst}\sim a^2\log\frac{16q}{x}=a^2(\frac{x}{2}+\frac{13 x^2}{64}+\frac{23 x^3}{192}+\frac{2701 x^4}{32768}+
\frac{5057 x^5}{81920}+\cdots)
\eea
which coincides with the expression given in part B3 of \cite{AGT}.
It would be interesting to extract the subleading $m$ corrections directly from the Seiberg-Witten curve too.

\subsection{$f=3$ antifundamentals}  
We have already seen how fruitful is the incorporation of AGT conjecture with Zamolodchikov's recursion relation. 
In order to get recursion relations also for less number (or even without) hypermultiplets one can decouple the extra 
hypermultiplets one after another by sending the masses to infinity.         
It is obvious from the Eq. (\ref{antifund}), (\ref{zinst}) that to decouple one of the hypermultiplets (say the one with mass $m_4$) 
one should renormalise the instanton parameter $x\rightarrow x/m_4$ and  go to the limit $m_4\rightarrow \infty$.
Similarly examining Zamolodchikov's recursion relation  (\ref{AlZamrecursion})  we see that there exists a smooth 
limit for the function $H$ at large $\mu_4$, provided one simultaneously redefines the parameter 
$q\rightarrow q/m_4=q/(\mu_4 \sqrt{\epsilon_1\epsilon_2})$. 
Indeed for large $\mu_4$ limit $R_{m,n}^{(4)}\sim \mu_4^{mn} R_{m,n}^{(3)}$. It remains to investigate the behaviour 
of the prefactors of the function $H$ in (\ref{ZversusH}). The analysis is elementary and boils down to expanding 
$\theta_3(q)$ up to first order: $\theta_3=1+2q+{\cal O}(q^2)$ and taking into account the relation (\ref{qexpansion}) 
between $q$ and $x$ (keeping first two terms is enough). Here is the result
\bea
Z_{inst}^{(3)}(a,m_i,x)= e^{-\frac{x}{4\epsilon _1 \epsilon _2} \left(\frac{x}{16}-\epsilon+2 m _1+2 m _2+2 m _3\right)}
H^{(3)}(\mu_i,\Delta,q),\label{Z3versusH}
\eea
where $q=\frac{x}{16\sqrt{\epsilon_1\epsilon_2}}$. Again at large $\Delta$  the function $H^{(3)}(\mu_i,\Delta,q)\sim 1$ and 
satisfies the recursion relation (\ref{AlZamrecursion}) with $R_{m,n}^{(4)}$ replaced by $R_{m,n}^{(3)}$ (see (\ref{Rmn})).

\subsection{Theories with $f=2,1$ or $0$}
It is straightforward to repeat the procedure of previous subsection and decouple more (or even all) hypermultiplets.
\begin{itemize}
\item $f=2$
\bea
Z_{inst}^{(2)}(a,m_1,m_2,x)= e^{-\frac{x}{2\epsilon _1 \epsilon _2} }
H^{(2)}(\mu_1,\mu_2,\Delta,q),\label{Z2versusH}
\eea
with $q=\frac{x}{16\epsilon_1\epsilon_2}$.
\item $f=1$
\bea
Z_{inst}^{(1)}(a,m_1,x)= H^{(1)}(\mu_1,\Delta,q),\label{Z1versusH}
\eea  
now with $q=\frac{x}{16(\epsilon_1\epsilon_2)^{3/2}}$.   
\item $f=0$
\bea
Z_{inst}(a,x)= H^{(0)}(\Delta,q),\label{Z0versusH}
\eea  
and $q=\frac{x}{16(\epsilon_1\epsilon_2)^{2}}$.  
\end{itemize}
The recursion relation for the pure ${\cal N}=2$ theory is especially simple and it is worth to rewrite it 
here in intrinsic terms:
\bea
Z_{inst}(a,x)=1-\sum_{m,n=1}^\infty\frac{x^{mn}{\cal R}_{m,n}}{4a^2-(m\epsilon_1+n\epsilon_2)^2}\,\, 
Z_{inst}((m\epsilon_1-n\epsilon_2)/2,x), \label{pureZrecursion}
\eea
where 
\bea
{\cal R}_{m,n}=2\prod^{\hspace{0.6cm}\prime}_{k,l}(k\epsilon_1+l\epsilon_2)^{-1}\,,\label{calRmn}
\eea
and the range of the product over $k,l$ is the same as in Eq. (\ref{Rmn}). 

It has been shown in recent papers \cite{Gaiottononconf,MarMirMor2} that when the the number of
fundamental hypermultiplets $f<4$ on CFT side one has irregular conformal blocks.

 
\section{Recursion relation for the case with adjoint hypermultiplet}\label{adjoint}
Encouraged with the success in the cases with fundamental hypermultiplets, it is natural to expect that 
a recursion relation of the same kind should exist also for the case of adjoint hypermultiplet or due to AGT conjecture
for the torus 1-point conformal block. In fact, explicit computation in first few orders of instanon expansion of the 
generalized partition function with adjoint and investigation of their large $a^2$ behaviour together with some intuition 
gained from the previous examples leads to the desired result. Define the function $H_{tor}$ by
\bea
Z_{inst}^{(adj)}(a,M,q)= \left[{\hat \eta}(q)\right]^{\frac{-2 \left(M-\epsilon _1\right)
 \left(M-\epsilon _2\right)}{\epsilon _1 \epsilon _2}}
H_{tor}(\mu,\Delta,q) \label{ZadjversusH}
\eea
where
\bea
{\hat \eta}(q)=\prod_{n=1}^\infty (1-q^n)
\eea
(we set $\mu=\frac{2 M-\epsilon}{2\sqrt{\epsilon_1\epsilon_2}}$, $\Delta=\frac{Q^2}{4}-\alpha^2$, $\alpha=\frac{a}{\sqrt{\epsilon_1\epsilon_2}}$
and instead of $x$ restore the conventional notation $q$ for instanton parameter). Then $H_{tor}=1+{\cal O}(\Delta)$ satisfies 
the relation
\bea
H_{tor}(\mu,\Delta,q)=1+\sum_{m,n=1}^\infty\frac{q^{mn}R_{m,n}^{(tor)}}{\Delta-\Delta_{m,n}} 
H_{tor}(\mu,\Delta_{m,n}+mn,q), \label{adjrecursion}
\eea
where
\bea
R_{m,n}^{(tor)}=R_{m,n}^{(4)}\label{Rmntor}
\eea
As earlier $R_{m,n}^{(4)}$ is given by Eq. (\ref{Rmn}) but the four parameters $\mu_i$ are specified as
\bea
\mu_1=\frac{M}{2\sqrt{\epsilon_1\epsilon_2}}\,; \quad
\mu_2=\frac{M+\epsilon_1}{2\sqrt{\epsilon_1\epsilon_2}}\,;\quad
\mu_3=\frac{M+\epsilon_2}{2\sqrt{\epsilon_1\epsilon_2}}\,;\quad
\mu_4=\frac{M+\epsilon_1+\epsilon_2}{2\sqrt{\epsilon_1\epsilon_2}}\label{mufundversusadj}
\eea 
I have checked the conjecture (\ref{ZadjversusH}), (\ref{adjrecursion}) up to 5 instantons.
The prefactor of $H_{tor}$ in (\ref{ZadjversusH}) defines the large $a^2$ behaviour of $Z_{adj}$ and hence
that of the prepotential
\bea
F^{SW}_{inst,adj}\sim M^2\log {\hat \eta}(q)
\eea
in agreement with the result derived from Seiberg-Witten curve (see e.g. \cite{FMPT}). 
 
Incorporating above results with AGT conjecture for adjoint hypermultiplet we find the equivalent 
recursion relation for torus 1-point conformal block 
\bea
{\cal F}_\alpha^\mu(q)=\left[ {\hat \eta}(q)\right]^{-1}H_{tor}(\mu,\Delta,q) \label{torversusH}
\eea
Observe that the recursion relation (\ref{AlZamrecursion}) together with asymptotic condition (\ref{asymptotic}) uniquely determines 
$H$ in terms of $R_{m,n}$. Thus the Eq's. (\ref{adjrecursion}), (\ref{Rmntor}) and (\ref{mufundversusadj}) lead to conclusion 
that
\bea
H_{tor}(\mu,\Delta,q)\label{HversusHtor}=H(\mu_1,\mu_2,\mu_3,\mu_4,\Delta,q)
\eea
provided the relations (\ref{mufundversusadj}) are hold.
This is a very exciting result: the torus 1-point block is closely related to specific sphere 4-point block on sphere 
or alternatively the generalized ${\cal N}=2$ SYM partition function with four hypermultiplets is related to 
the partition function with adjoint hypermultiplet. On CFT side e.g. one gets  (in the reminder of this section 
the relation (\ref{xversusq}) between parameters $q$ and $x$ is always assumed)
\bea
{\cal F}_\alpha^\mu(q)&=&\left[ {\hat \eta}(q)\right]^{-1}
\left(16q\right)^{\alpha^2}x^{-Q^2/4+\Delta_1+\Delta_2}\nonumber\\
&&\times (1-x)^{-Q^2/4+\Delta_1+\Delta_3}\theta_3(q)^{-3Q^2+4(\Delta_1+\Delta_2+\Delta_3+\Delta_4)}{\cal F}(\Delta_i,\Delta,x)
\eea
Owing to already mentioned reach symmetry of the function ${\cal F}$ for given dimensions 
$\Delta=Q^2/4-\alpha^2 $ and  $\Delta_\mu=Q^2/4-\mu^2 $ the choice of $\Delta_i=Q^2/4-\lambda_i^2$ is not unique. 
The essentially different choices are
\bea
\lambda_1=\frac{\mu}{2}\,;\quad \lambda_2=\frac{Q}{4}\,;\quad 
\lambda_3=\frac{\mu}{2}\,;\quad\lambda_4=\frac{1}{4}\sqrt{Q^2-4}
\eea 
and somewhat more sophisticated 
\bea
\lambda_1&=&\frac{\mu}{2}-\frac{1}{8}\left(Q+\sqrt{Q^2-4}\right)\,;\quad \lambda_2=\frac{1}{8}\left(Q-\sqrt{Q^2-4}\right)\,;\nonumber \\
\lambda_3&=&\frac{\mu}{2}+\frac{1}{8}\left(Q+\sqrt{Q^2-4}\right)\,;\,\,\,\quad\lambda_4=\frac{1}{8}\left(Q-\sqrt{Q^2-4}\right).
\eea 
Since the calculation of the torus 1-point function ${\cal F}_\alpha^\mu(q)$ from first principles of 
CFT is less familiar for reader's convenience I sketch the procedure in the appendix.

The Eq.s (\ref{ZversusF}), (\ref{ZadjversusH}) and (\ref{torversusH}) straightforwardly lead to analogues 
(and by no means less surprising) relation among SYM partition function with four hypermultiplets and  the partition 
function with adjoint hypermultiplet 
\bea 
Z_{inst}^{(adj)}(a,M,q)&=& \left[{\hat \eta}(q)\right]^{\frac{-2 \left(M-\epsilon _1\right)
 \left(M-\epsilon _2\right)}{\epsilon _1 \epsilon _2}}
\left(\frac{x}{16q}\right)^{-\frac{a^2}{\epsilon_1\epsilon_2}} 
(1-x)^{-\frac{1}{4\epsilon_1\epsilon_2}(\epsilon-\sum_{i=1}^4m_i)^2}\nonumber\\
&\times & 
\left[\theta_3(q)\right]^{-\frac{1}{\epsilon _1\epsilon _2}\sum _{i=1}^4 \left(m _i^2+
\left(\epsilon-m _i\right)^2-3 \epsilon^2/4\right)}
Z_{inst}^{(4)}(a,m_i,x)
\eea
Here the masses $m_i=\sqrt{\epsilon_1\epsilon_2}\,\mu_i$ (up to permutations) are given by Eq. (\ref{mufundversusadj}).

\section*{Acknowledgements}
It is a pleasure to thank H.~Babujian, M.~Bianchi, R.~Flume, F.~Fucito, R.~Manvelyan, J.-F.~Morales and M.~Samsonyan for
interesting discussions. I am especially thankful to A.Belavin who some time ago explained me 
Zamolodchikov's recursion relations in full details.
This work was partly supported by the Institutional Partnership grant of Humboldt
foundation of Germany, EC FP7 Programme Marie Curie Grant Agreement PIIF-GA-2008-221571,
the Advanced Grant n.226455, “Supersymmetry, Quantum Gravity and Gauge 
Fields (SUPERFIELDS) and by the Italian MIUR-PRIN contract 20075ATT78.

\section*{Note added}
After the first version of this paper appeared in arXiv Vl.~Fateev, A.~Litvinov and S.~Ribault kindly informed me about
the work \cite{Fateev} where a relation between the 1-point correlation function on the torus and 
a specific 4-point correlation function on sphere is established. Contrary to the one, suggested in this article,
that relation holds for different (though simply related) values of the central charge on the sphere and torus. 
To avoid confusion let me note also that in present paper the modular parameter of the torus is related to the 
parameter $q$ introduced in section \ref{qrecursion}. via $q=\exp (2\pi i\tau_{tor})$, \ie $\tau_{tor}=\tau/2$ 
with $\tau$ given by Eq. (\ref{tau}) while the modular parameter of the Ref. \cite{Fateev} is equal to  $\tau$.  
Naturally the relation of Ref. \cite{Fateev} between correlation functions boils down to a certain relation among
conformal blocks. The condition that this relation and the one conjectured in present paper are compatible 
is equivalent to the following non-trivial identity  
(see Eq. (\ref{H}) for definition of the function $H$ and the Eq.'s (\ref{mu-lambda}), (\ref{dimensions}) which relate its arguments  
$\mu_i$ to the dimensions of external fields):
\bea
H_{\tilde{b}}(\tilde{\mu}_i,\tilde{\Delta} ,q^2)=H_b(\mu_i,\Delta ,q)\nonumber
\eea 
where the dependence on the parameter $b$ specifying the central charge is indicated explicitly and the remaining 
parameters are specified as (the parameters established in present paper are on the first line while those on second line 
are found in \cite{Fateev}):  
\bea
&&\tilde{\mu}_1=\frac{\tilde{\mu}}{2}+\frac{\tilde{b}}{4}+\frac{1}{4\tilde{b}}\,;\quad 
\tilde{\mu}_2=\frac{\tilde{\mu}}{2}+\frac{3\tilde{b}}{4}+\frac{1}{4\tilde{b}} \,;\quad 
\tilde{\mu}_3=\frac{\tilde{\mu}}{2}+\frac{\tilde{b}}{4}+\frac{3}{4\tilde{b}} \,;\quad 
\tilde{\mu}_4=\frac{\tilde{\mu}}{2}+\frac{3\tilde{b}}{4}+\frac{3}{4\tilde{b}}\nonumber \\
&&\mu_1=\mu+\frac{b}{4}+\frac{1}{2b}\,;\quad \mu_2=\mu+\frac{3b}{4}+\frac{1}{2b}\,;\quad 
\mu_3=b+\frac{1}{2b}\,;\quad \mu_4=\frac{b}{2}+\frac{1}{2b}\nonumber 
\eea     
and 
\[
\tilde{b}=\frac{b}{\sqrt{2}}\,;\quad \tilde{\mu}=\sqrt{2}\mu\,;\quad \tilde{\Delta }=\frac{(\tilde{b}+\frac{1}{\tilde{b}})^2}{4}-
\tilde{\alpha}^2 \,;\quad \Delta=\frac{(b+\frac{1}{b})^2}{4}-\alpha^2\,;\quad \tilde{\alpha}=\frac{\alpha}{\sqrt{2}}
\]
It should be possible to find a mathematical proof for this identity based on Zamolodchikov recursion relation (\ref{AlZamrecursion}). 
Direct calculation using the recursion relation (\ref{AlZamrecursion}) confirms that the identity indeed holds up to high orders in $q$.  

\appendix

\section{Torus 1-point block}\label{torusblock}
\setcounter{equation}{0}

Below is presented the calculation of the one-point conformal block on torus \\
$tr_\alpha \left(q^{L_0-c/12} \phi_\mu\right)$ up to level 2 
(in principle the computation can be carried out up to arbitrary level). Start with (chiral) OPE
\bea
\phi_\mu(x)\phi_\alpha(0)&=&\sum_{Y} x^{-\Delta_\mu+|Y|}\beta_{\mu\alpha}^{\alpha Y}L_{-Y}\phi_{\alpha}(0)= 
\nonumber\\ 
&&x^{-\Delta_\mu}(1+x\beta^{1}L_{-1}+x^2(\beta^{11}L_{-1}^2+\beta^2L_{-2})+\cdots)\phi_\alpha(0) \label{OPE of 2 primaries}
\eea
where for the partition $Y=\{k_1\geq k_2 \geq \cdots \geq 0\}$, $|Y|=k_1+ k_2 + \cdots $, 
\bea
L_{-Y}\equiv L_{-k_1}L_{-k_2}\cdots
\eea
and $L_k$ are the standard Virasoro generators. It is well known that the coefficients $\beta$ in principle could be 
calculated level by level using conformal symmetry 
\cite{BPZ}. 
In particular the first few coefficients $\beta$ explicitly presented in (\ref{OPE of 2 primaries}) are 
\bea
\beta^{\{1\}}&=&\frac{\Delta_\mu}{2 \Delta}\nonumber\\
\beta^{\{11\}}&=&\frac{\Delta _\mu \left(c-16 \Delta +(c+8 \Delta ) \Delta _\mu\right)}{4 \Delta  (c+2 c \Delta +2 \Delta  (-5+8 \Delta ))}, \nonumber \\ 
\beta^{\{2\}}_2&=& \frac{\left(1+8 \Delta -3 \Delta _\mu\right) \Delta _\mu}{c+2 c \Delta +2 \Delta  (-5+8 \Delta )}
\eea
To calculate the trace the diagonal matrix elements of the OPE of the primary field $\phi_M$ with the descendants of the 
field $\phi_\alpha$ are needed. Using the commutation relation 
\be 
[L_n,\phi_\mu(x)]=x^n(x\partial+(1+n)\Delta_\mu)\phi_\mu(x)
\ee  
one easily finds
\bea 
\phi_\mu(x)L_{-1}\phi_\alpha(0)&=&x^{-\Delta_\mu}(x^{-1}\Delta_\mu+(1+\beta^1(\Delta_\mu-1))L_{-1}+\cdots)\phi_\alpha(0)\nonumber\\
\phi_\mu(x)L_{-2}\phi_\alpha(0)&=&x^{-\Delta_\mu}(x^{-1}\Delta_\mu+(1+\beta^1(\Delta_\mu-1))L_{-1}+\cdots)\phi_\alpha(0)\nonumber\\
\phi_\mu(x)L_{-1}^2\phi_\alpha(0)&=&x^{-\Delta_\mu}(x^{-2}(\Delta_\mu-1)(\Delta_\mu+1)(\beta^{11}L_{-1}^2+\beta^2L_{-2})\nonumber\\ 
&+&x^{-1}2(\Delta_\mu-1)L_{-1}+L_{-1}^2+\cdots)\phi_\alpha(0)\label{OPE with descendant}
\eea
It remains to read off the diagonal matrix elements from (\ref{OPE with descendant})  to get the torus one point block 
\bea
{\cal F}_\alpha^\mu(q)\equiv q^{-\Delta_\alpha +\frac{c}{12}}
tr_\alpha \left(q^{L_0-c/12} \phi_\mu(1)\right)=1+\left(1+\left(\Delta _\mu-1\right) \beta^{\{1\}}\right) q \hspace{2.5cm}\nonumber \\
+\left(\left(\Delta _\mu-1\right) \left(\Delta _\mu-2\right) \beta^{\{11\}}
 2 \left(\Delta _\mu-1\right) \beta^{\{1\}}+2 \left(\Delta _\mu-1\right) \beta^{\{2\}}+2 \right)  q^2+\cdots \quad
\eea
It is easy to check that this result is consistent with recursion relation (\ref{adjrecursion}), (\ref{torversusH}).

 \bibliographystyle{abe}

\end{document}